\documentclass{appolb}
\usepackage{graphicx}
\usepackage{booktabs}
\usepackage[italic]{hepnames}
\usepackage{amsfonts}
\usepackage{amsthm}
\usepackage{bm}
\usepackage{slashed}
\usepackage{dsfont}
\usepackage{tikz}
\usepackage[dvipsnames]{xcolor}
\usepackage[shortlabels]{enumitem}
\usepackage{xspace}
\usepackage{siunitx}
\usepackage{bookmark}
\usepackage{bbm}
\usepackage{subcaption}

\def\1eq#1{Eq.\nobreak\thinspace(\ref{#1})}

\def\2eqs#1#2{Eqs.\nobreak\thinspace(\ref{#1}) and\nobreak\thinspace(\ref{#2})}
\def\3eqs#1#2#3{Eqs.\nobreak\thinspace(\ref{#1}),\nobreak\thinspace(\ref{#2}) and\nobreak\thinspace(\ref{#3})}

\def\fig#1{\hyperref[#1]{Fig.\nobreak\thinspace\ref*{#1}}}
\def\figA#1{\hyperref[#1]{Fig.\nobreak\thinspace\ref*{#1}A}}
\def\figB#1{\hyperref[#1]{Fig.\nobreak\thinspace\ref*{#1}B}}
\def\figC#1{\hyperref[#1]{Fig.\nobreak\thinspace\ref*{#1}C}}
\def\tab#1{\hyperref[#1]{Tab.\nobreak\thinspace\ref*{#1}}}

\def\sect#1{\hyperref[#1]{Sec.\nobreak\thinspace\ref*{#1}}}
\def\appref#1{\hyperref[#1]{App.\nobreak\thinspace\ref*{#1}}}

\newcommand{\be}{\begin{equation}}
\newcommand{\ee}{\end{equation}}

\newcommand{\bea}{\begin{eqnarray}}
\newcommand{\eea}{\end{eqnarray}}

\begin{document} 
\title{Electromagnetic form factors of 
heavy-light
\\ pseudoscalar mesons.%
\thanks{Presented by A.S.M. at the Excited QCD Workshop 2026, Granada, Spain}%
}
\author{A.S.~Miramontes and J.~Papavassiliou
\address{\mbox{Department of Theoretical Physics and IFIC, University of Valencia and CSIC}, E-46100, Valencia, Spain}
\\[3mm]
J.M.~Pawlowski
\address{\mbox{Institut f\"ur Theoretische Physik, Universit\"at Heidelberg}, Philosophenweg 16, Heidelberg, 69120, Germany}
\address{\mbox{ExtreMe Matter Institute EMMI, GSI, Planckstrasse 1, Darmstadt, 64291, Germany}}}

\maketitle
\begin{abstract}
We report calculations of space-like electromagnetic form factors and charge radii of pseudoscalar mesons, covering both light and heavy-light flavour sectors within a flavour-dependent Bethe-Salpeter framework.
\end{abstract}
  
\section{Introduction}
Electromagnetic form factors (EFF) of pseudoscalar mesons are central observables for elucidating hadron structure in terms of quark and gluon dynamics. Considerable progress has been achieved over the years in mapping these form factors, driven by both improved experimental determinations and increasingly refined theoretical treatments. On the theory side, functional approaches based on the Bethe-Salpeter equations (BSEs) and Schwinger-Dyson equations (SDEs) provide a self-consistent framework for addressing bound state properties, and have been widely employed for light meson form factors and spectroscopy \cite{Eichmann:2016yit, Sanchis-Alepuz:2017jjd}. By contrast, extending the same level of control to heavy-light systems introduces additional complications, mainly due to the strong flavour asymmetry between the light and heavy quark entering into the bound state equations.
In this contribution, we report on recent SDE/BSE computations of space-like electromagnetic form factors for pseudoscalar mesons across different flavours \cite{Miramontes:2025vzb}, ranging from $\pi^{\pm}$ and $K^{\pm}$ to $D, D_s, B$ and $B_s$.

\section{Formalism}
\begin{figure*}[t]
\centerline{%
\includegraphics[width=0.8\textwidth]{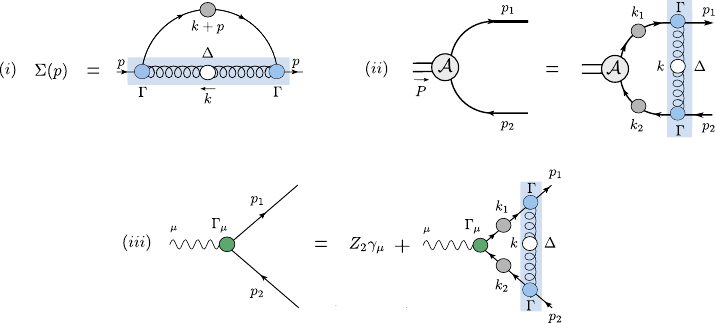}}
\caption{Diagrammatic representation of the main dynamical components: (i) is the quark self-energy; (ii) the BSE for the mesonic amplitude; and (iii) the SDE of the quark-photon vertex.}
\label{fig:truncation}     
\end{figure*}

In this section, we briefly summarize the ingredients involved in the calculation of pseudoscalar electromagnetic form factors. The central dynamical components are illustrated in \fig{fig:truncation}: (i) the quark self energy, (ii) the meson BS amplitude, and (iii) the quark-photon vertex.
These quantities serve as the building blocks for evaluating the hadronic matrix element. Specifically, the electromagnetic current between pseudoscalar meson states is written as
\begin{equation}
J_\mu(p_{\rm av},q)=\langle s(p_f)\,|\,j_\mu(0)\,|\,s(p_i)\rangle\,,
\qquad
j_\mu(x)=\bar\psi(x)\gamma_\mu\psi(x)\,,
\end{equation}
with $q=p_f-p_i$ and $p_{\rm av}=(p_f+p_i)/2$. The corresponding form factor is defined through
\begin{equation}
J^\mu(p_{\rm av},q)=2\,p_{\rm av}^{\mu}F_s(q^2)\,.
\end{equation}
In the impulse approximation, the photon couples to the valence quarks inside the meson such that
\begin{equation}
J_\mu(p_{\rm av},q)=\int_k
\bar{\mathcal{A}}(k_f,p_f)\,
S(\ell_+)\,
\Gamma_\mu(k_+,q)\,
S(\ell_-)\,
\mathcal{A}(k_i,p_i)\,
S(k_-)\,,
\label{eq:current}
\end{equation}
where $\mathcal{A}$ and $\bar{\mathcal{A}}$ denote the BS amplitude and its charge-conjugated counterpart, $S$ is the dressed quark propagator, and $\Gamma_\mu$ is the quark-photon vertex. The internal momenta are
\begin{equation}
k_\pm=k\pm\eta p_{\rm av},\qquad
\ell_\pm=k_+\pm q/2,\qquad
s_\pm=k_-\pm q/2\,,
\label{eq:kinematics}
\end{equation}
with $\eta$ controlling the momentum partition between the quark and antiquark. In practice, $\eta$ is chosen for each flavour combination to avoid propagator singularities and improve numerical stability, particularly in heavy-light systems. 

The full EFF combines both valence constituent contributions, weighted by their electric charges:
\begin{equation}
F_s(q^2)=e_q\,F_q(q^2)+e_{\bar q}\,F_{\bar q}(q^2)\,.
\end{equation}

\begin{figure*}[t]
\centerline{%
\includegraphics[width=0.925\textwidth]{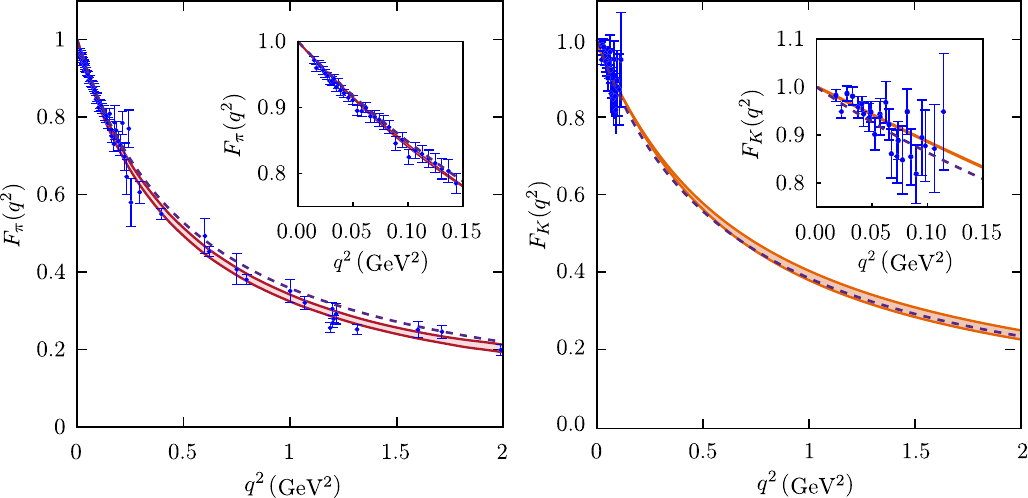}}
\caption{Pion (left) and kaon (right) electromagnetic form factors for a space-like photon. Solid lines and bands show our results and associated $\tilde\alpha_T(q^2)$ uncertainties; dashed lines correspond to Ref.~\cite{Xu:2024fun}. Experimental data from Refs.~\cite{JeffersonLabFpi-2:2006ysh, JeffersonLabFpi:2000nlc, NA7:1986vav, Dally:1980dj}.
}
\label{fig:pionFF}
\end{figure*}

The ingredients entering \1eq{eq:current} are computed in a flavour-dependent interaction. The corresponding interaction is written as
\begin{equation}
I_{ff'}(q^2)=\tilde\alpha_T(q^2)\,A_f(q^2)\,A_{f'}(q^2)\,,
\end{equation}
where $\tilde\alpha_T(q^2)$ includes additional vertex contributions beyond the standard Taylor coupling \cite{Gao:2024gdj}. The associated kernel reads:
\begin{equation}
D^{ff'}_{\mu\nu}(k)=4\pi\,D^0_{\mu\nu}(k)\,I_{ff'}(k^2)\,.
\end{equation}

The dressed quark propagator is decomposed as
$S_f^{-1}(p)=i\!\not\!p\,A_f(p^2)+B_f(p^2)\,$,
with the mass function $M_f(p^2)=B_f(p^2)/A_f(p^2)$. It is obtained from the gap equation
\begin{equation}
S_f^{-1}(p)=Z_2\,(i\!\not\!p+m_R)+C_F\int_k \gamma_\alpha S_f(k+p)\gamma_\beta D^{ff}_{\alpha\beta}(k)\,,
\end{equation}
where $Z_2$ is the renormalization constant, $C_F=4/3$ and $m_R$ is the renormalized current quark mass.
\begin{figure*}[t]
\centerline{%
\includegraphics[width=0.83\textwidth]{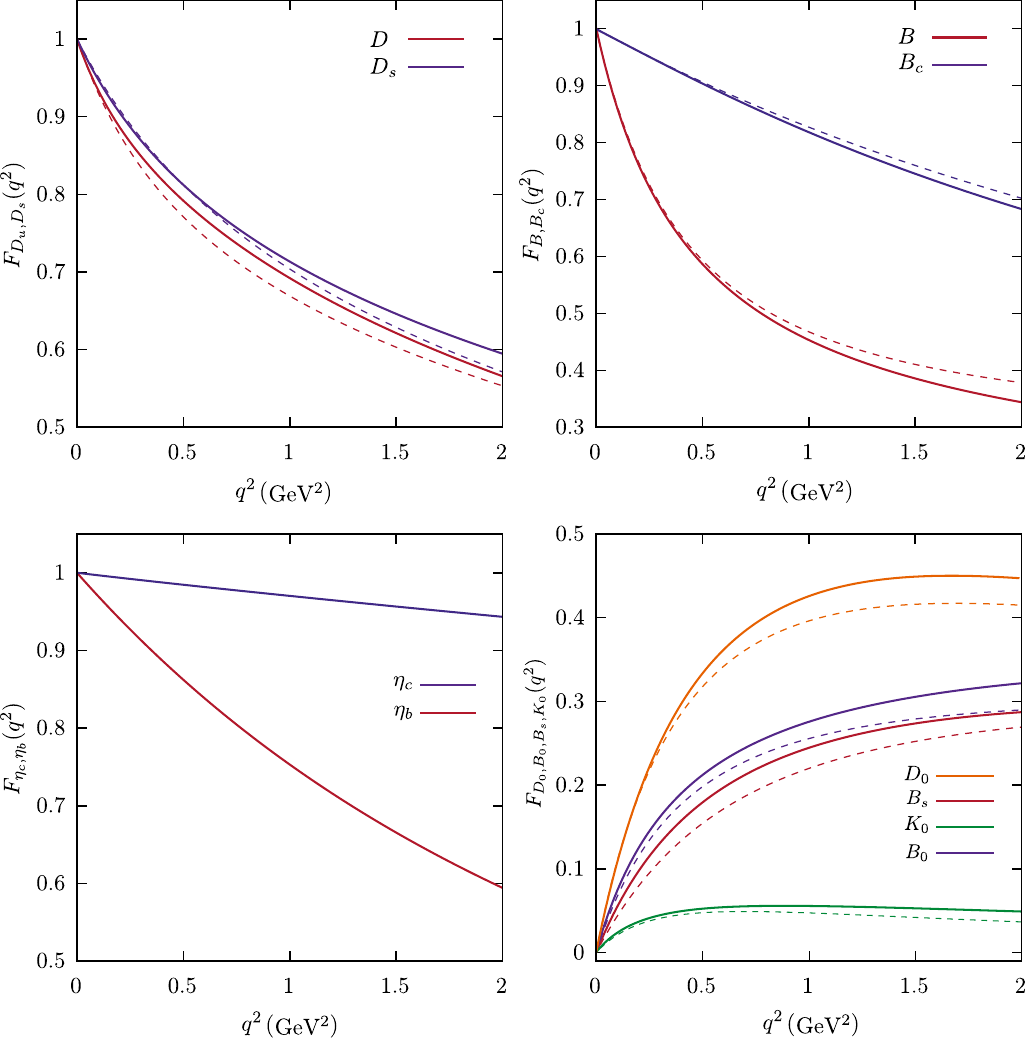}}
\caption{Meson electromagnetic form factors for a space-like photon. Solid lines show our results, while dashed lines correspond to Ref.~\cite{Xu:2024fun}.}
\label{fig:HLFF}     
\end{figure*}
On the other hand, the pseudoscalar BSA satisfies
\begin{equation}
\mathcal{A}_{ff'}(p_1,p_2)=
-\int_k \gamma_\mu S_f(k_1)\mathcal{A}_{ff'}(k_1,k_2)S_{f'}(k_2)\gamma_\nu D^{ff'}_{\mu\nu}(k)\,,
\end{equation}
while the quark-photon vertex obeys
\begin{equation}
\Gamma^f_\mu(p_1,p_2)=
Z_2\gamma_\mu-C_F\int_k \gamma_\alpha S_f(k_1)\Gamma^f_\mu(k_1,k_2)S_f(k_2)\gamma_\beta D^{ff}_{\alpha\beta}(k)\,.
\end{equation}
Its longitudinal part satisfies the vector Ward-Takahashi identity (VWTI), while the transverse sector accounts for additional nonperturbative structures \cite{Miramontes:2019mco}.

\section{Results}
\begin{table*}[t]
\centering
\scriptsize
\begin{tabular}{lccccc}
\hline
Meson & This work & Exp.\cite{ParticleDataGroup:2022pth} & LQCD \cite{Wang:2020nbf, Can:2012tx,Vujmilovic:2025czt} & RL~\cite{Xu:2024fun} & LF~\cite{Xu:2025ntz} \\
\hline
$\pi^+$   & 0.656(5)   & 0.659(4)   & 0.656(11) & 0.646    & 0.668(35) \\
$K^+$     & 0.568(4)   & 0.560(30)  & --         & 0.608    & 0.610(20) \\
$K^0$     & 0.270\,$i$   & 0.277(18)\,$i$ & --       & 0.253\,$i$ & 0.302(22)\,$i$ \\
$D$       & 0.428      & --         & 0.450(24) & 0.435    & 0.411(15) \\
$D^0$     & 0.542\,$i$   & --         & --         & 0.556\,$i$ & 0.534(26)\,$i$ \\
$D_s$     & 0.368      & --         & 0.465(57) & 0.352    & 0.301(11) \\
$B$       & 0.631      & --         & 0.692(21)         & 0.619    & 0.564(22) \\
$B^0$     & 0.442\,$i$   & --         & --         & 0.435\,$i$ & 0.396(16)\,$i$ \\
$B_s$     & 0.330\,$i$   & --         & --         & 0.337\,$i$ & 0.281(13)\,$i$ \\
$B_c$     & 0.213      & --         & --         & 0.219    & 0.189(10) \\
$\eta_c$  & 0.267      & --         & --         & --       & -- \\
$\eta_b$  & 0.082      & --         & --         & --       & -- \\
\hline
\end{tabular}
\caption{Charge radii in fm. For neutral states with negative squared radii, the results are written with an explicit factor of $i$. The $\eta_c$ and $\eta_b$ entries correspond to single-quark contributions and are included as theoretical benchmarks.}
\label{tab:radii}
\end{table*}

In this section we summarize our results for space-like EFF $F(Q^2)$ of pseudoscalar mesons. All calculations employ the input current-quark masses fixed at $\mu=4.3$\,GeV: $m_{u/d}=0.005$\,GeV, $m_s=0.094$\,GeV, $m_c=1.1$\,GeV, $m_b=3.5$\,GeV, with $\eta$-values in \1eq{eq:kinematics} tuned per flavour content for numerical stability. For further details, see discussion in  \cite{Miramontes:2025vzb}.

In \fig{fig:pionFF} we display our computed pion and kaon electromagnetic form factors. Solid lines and bands denote our central results and the uncertainty from $\tilde\alpha_T(q^2)$ variations, respectively. The pion result is in excellent agreement with experiment, while for the kaon, our prediction matches the available low-$Q^2$ data within uncertainties.

For heavy-light systems (\fig{fig:HLFF}), we present results for the $D$, $D_s$, and $B/B_c$ form factors, together with single-quark contributions for the $\eta_c$ and $\eta_b$, which, although not physical observables, provide useful benchmarks for heavy-heavy systems. The $D$ and $D_s$ form factors show similar $Q^2$ dependence, whereas the $B$-meson results are more compact due to the heavier bottom quark. We also show the neutral-sector form factors for $K^0$, $D^0$, $B^0$, and $B_s^0$, which vanish at $Q^2=0$ as required by charge conservation and the VWTI, while retaining nontrivial sensitivity to the flavour structure of the corresponding bound states.

Finally, the charge radii are extracted via $\langle r^2\rangle=-6\,dF/dQ^2|_{Q^2=0}$. Table~\ref{tab:radii} compares our results to experiments, lattice QCD (LQCD), rainbow-ladder (RL) and Light-Front (LF) model.

\section{Conclusions}
We have presented a BSE study of space-like electromagnetic form factors for light and heavy-light pseudoscalar mesons, based on a flavour-dependent interaction kernel. While the pion and kaon results reproduce the available experimental data, our predictions for the form factors and charge radii of heavy-light mesons are in overall agreement with current theoretical approaches.

\section*{Acknowledgments}
The work of A.S.M. and J.P. is funded by the Spanish MICINN grants PID2020-113334GB-I00 and PID2023-151418NB-I00, the Generalitat Valenciana grant CIPROM/2022/66, and CEX2023-001292-S by MCIU/AEI. J.M.P. is funded by the Deutsche Forschungsgemeinschaft (DFG, German Research Foundation) under Germany’s Excellence Strategy EXC 2181/1 - 390900948 (the Heidelberg STRUCTURES Excellence Cluster) and the Collaborative Research Centre SFB 1225 - 273811115 (ISOQUANT).

\bibliographystyle{polonica}
\bibliography{bibliography.bib}

\end{document}